\documentclass[twocolumn,showpacs,preprintnumbers,amsmath,amssymb]{revtex4}
\usepackage{graphicx}% Include figure files
\usepackage{dcolumn}% Align table columns on decimal point
\usepackage{bm}% bold math
\usepackage{psfig}

\begin{document}

\preprint{APS/123-QED}

\title{Spin Splitting in GaAs (100) Two-Dimensional Holes Revisited}

\author{B.~Habib, E.~Tutuc, S.~Melinte, M.~Shayegan, D.~Wasserman, S.~A.~Lyon}
%\email[Corresponding author, email address:
%]{bhabib@princeton.edu}
%\author{E.~Tutuc, S.~Melinte, M.~Shayegan, D.~Wasserman, S.~A.~Lyon}

\affiliation{Department of Electrical Engineering, Princeton
University, Princeton, NJ 08544, USA}

\author{R.~Winkler}
\affiliation {Institut f\"{u}r Festk\"orperphysik,
Universit\"{a}t Hannover, Appelstr.~2, D-30167
Hannover, Germany}

\date{\today}

\begin{abstract}
We measured Shubnikov-de Haas (SdH) oscillations in GaAs (100)
two-dimensional holes to determine the inversion asymmetry-induced
spin splitting. The Fourier spectrum of the SdH oscillations
contains two peaks, at frequencies $f_-$ and $f_+$, that
correspond to the hole densities of the two spin subbands and a
peak, at frequency $f_\mathrm{tot}$, corresponding to the total
hole density. In addition, the spectrum exhibits an anomalous peak
at $f_\mathrm{tot}/2$. We also determined the effective masses of
the two spin subbands by finding the inverse transform of the
Fourier spectrum in the vicinity of $f_-$ and $f_+$, and then
analyzing the temperature dependence of the SdH oscillations for
each subband. We discuss our results in light of self-consistent
calculations and previous experiments.
\end{abstract}

\pacs{Valid PACS appear here}% PACS, the Physics and Astronomy
                             % Classification Scheme.
%\keywords{Suggested keywords}%Use showkeys class option if keyword
                              %display desired
\maketitle

%%%%%%%%%%%%%%%%%%%%%%%%%%%%%%%%%%%%%%%%%%%%%%%%%%%%%%%%%%%%%%%%%%
% \section{\label{sec:level1}Introduction}

\begin{figure*}
\centering
\includegraphics {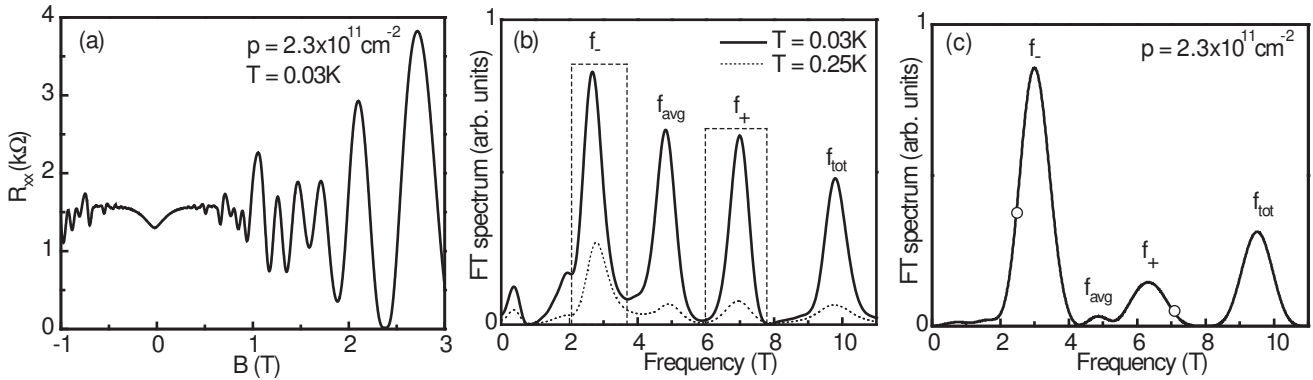}
\caption{\label{fig:fft}(a) Observed Shubnikov-de Haas
oscillations for a 2D hole system confined to a (100) GaAs/AlGaAs
heterojunction. (b) The Fourier spectrum of the oscillations (in
the range $0.3T\leq B\geq 3T$) at two different temperatures. The
dashed boxes show the square windows applied to isolate the $f_-$
and $f_+$ peaks before obtaining the inverse Fourier transforms
which are shown in the insets in Fig.~\ref{fig:ifft}. (c) Fourier
spectrum of calculated magneto-oscillations. The open circles
indicate the expected peak positions according to the calculated
spin subband densities at zero magnetic field (see text).}
\end{figure*}

In a solid that lacks inversion symmetry, the spin-orbit (SO)
interaction leads to a lifting of the spin degeneracy of the
energy bands, even in the absence of an applied magnetic field. In
such a solid, the energy bands at finite wave vectors are split
into two spin subbands with different energy surfaces,
populations, and effective masses. This so-called zero-field spin
splitting is particularly important for semiconductor holes
because they typically have much smaller kinetic energies than
electrons. The problem of inversion asymmetry-induced spin
splitting in two-dimensional (2D) carrier systems in semiconductor
heterojunctions and quantum wells \cite{Stormer, Eisenstein,
Wieck} has become of renewed interest recently \cite{Winkler03}
because of the possible use of such systems in realizing
spintronic devices such as a spin field-effect transistor
\cite{Prinz, Datta}, and for studying fundamental phenomena such
as the spin Berry phase \cite{Morpurgo, Yau}.

GaAs 2D holes in modulation-doped GaAs/AlGaAs heterostructures
\cite{Stormer, Eisenstein} provide an attractive system for these
studies as they can have high low-temperature mobilities and a
strong SO interaction that is tunable via the application of gate
bias \cite{Lu, Papadakis}. Recent work in these systems has
concentrated primarily on 2D holes that are fabricated, via
molecular beam epitaxy (MBE), on GaAs (311)A substrates.  The main
advantage of using (311)A substrates, rather than the more
commonly used (100) substrates, is that one can use Si as a
$p$-type dopant in the MBE growth.  The standard $p$-type dopant
for MBE growth on GaAs (100) is Be. Unlike Si, however, Be has a
tendency to diffuse and migrate under typical MBE growth
conditions and, as a result, fabrication of high-quality GaAs
(100) 2D holes has been challenging.  On the other hand, because
of a higher symmetry, the band structure for (100) GaAs 2D holes
is simpler than for (311)A \cite{Winkler03}. It is therefore
desirable to study the problem of spin splitting in GaAs (100) 2D
hole systems. We have been able to fabricate such 2D systems with
reasonably high mobilities, and report here our measurements of
their spin splitting and of the hole effective mass, and the
comparison of the experimental data with the results of
state-of-the-art energy band calculations \cite{Winkler03,
Winkler93}.

The holes in GaAs (100) were in fact the first 2D system in which
the inversion asymmetry-induced spin splitting was observed
experimentally using SdH oscillations \cite{Stormer, Eisenstein}.
The oscillations in 2D holes confined to GaAs/AlGaAs
heterojunctions revealed a beating pattern, indicating the
presence of two spin subbands with different hole densities
\cite{Eisenstein}. The spin splitting was inferred from a change
in the frequency of the oscillations above a certain magnetic
field. The temperature dependence of the amplitude of the
oscillations along with the assumption that the spin subbands are
parabolic was used to deduce their masses \cite{Eisenstein}.
Although the results of these experiments were in qualitative
agreement with the calculations that followed, some questions
remain, particularly regarding the effective mass values
\cite{Broido, Ando85, Ekenberg, Bangert}. In our study, we revisit
the spin splitting in GaAs (100) 2D holes via a careful analysis
of the beating pattern in SdH oscillations. We employ Fourier
transform (FT) techniques to determine the temperature dependence
of the amplitude of the SdH oscillations and the effective masses
for both the lighter (HHl) and heavier (HHh) heavy-hole spin
subbands independently. Our method has the advantage that it does
not require assumptions regarding the band structure, namely, the
parabolicity of the spin subbands for determining their effective
masses. The masses we find are in good agreement with the results
of subband calculations. The data and calculations also provide
clear evidence for the strong nonparabolicity of the HHh band.

%%%%%%%%%%%%%%%%%%%%%%%%%%%%%%%%%%%%%%%%%%%%%%%%%%%%%%%%%%%%%%%%%%
% \section{Experimental Results}

Our samples were grown on GaAs (100) substrates by MBE and contain
modulation-doped 2D hole systems that are confined to either a
20~nm-wide GaAs square quantum well or a GaAs/AlGaAs
heterojunction. The square well is flanked on each side by undoped
Al$_{0.3}$Ga$_{0.7}$As spacer layers. On the front (surface) side,
a Be-doped layer (Be concentration of $2.6 \times
10^{18}$~cm$^{-3}$) of Al$_{0.3}$Ga$_{0.7}$ follows the 21~nm
thick Al$_{0.3}$Ga$_{0.7}$ spacer layer. The
Al$_{0.3}$Ga$_{0.7}$As/GaAs interface in the heterojunction sample
is separated from a 16~nm thick Be-doped Al$_{0.3}$Ga$_{0.7}$As
layer (Be concentration of $3.5 \times 10^{18}$~cm$^{-3}$) by a
25~nm Al$_{0.3}$Ga$_{0.7}$As spacer layer. In order to reduce Be
diffusion and migration during the MBE growth, and hence increase
the quality of the samples, the substrate temperature was lowered
to 550$^\circ$C from 640$^\circ$C before doping. In addition, for
the square well sample, the doping was done on the front side of
the well only unlike previous samples, which had doping on both
the front and the substrate sides. We fabricated Hall bar patterns
on all the samples and used In/Zn alloyed at 440$^\circ$C for the
ohmic contacts. Metal gates were deposited on the front and the
back of the samples to control the density. The typical low
temperature mobility for the square well sample is $2.6 \times
10^5$~cm$^2$/Vs at a 2D hole density of $p = 1.7 \times
10^{11}$~cm$^{-2}$ and for the heterojunction sample it is $7.7
\times 10^4$~cm$^2$/Vs at $p = 2.3 \times 10^{11}$~cm$^{-2}$.
Longitudinal ($R_{xx}$) and transverse ($R_{xy}$)
magneto-resistances were measured, as a function of the
perpendicular magnetic field ($B$), at $T \approx 30$~mK via a
standard low frequency lock-in technique.

Figure~\ref{fig:fft}(a) shows the low-field SdH oscillations for
the heterojunction sample. The beating pattern in the oscillations
indicates that the two spin subbands are nondegenerate. The FT
spectrum of the oscillations, shown in Fig.~\ref{fig:fft}(b),
exhibits four dominant peaks at frequencies $f_-$,
$f_\mathrm{avg}$, $f_+$, and $f_\mathrm{tot}$, with the relation
$f_\mathrm{tot} = f_+ + f_- = 2f_\mathrm{avg}$. The
$f_\mathrm{tot}$ frequency, when multiplied by $e/h$, matches well
the total 2D hole density deduced from the Hall resistance ($e$ is
the electron charge and $h$ is Planck's constant). The two peaks
at $f_-$ and $f_+$ correspond to the SdH oscillations of the holes
in individual spin subbands although, as discussed below, their
positions times $e/h$ do not exactly give the spin subband
densities. As we also discuss later in the text, the presence of a
peak at $f_\mathrm{avg}$ is related to this anomalous behavior. We
note that the $f_\mathrm{avg}$ peak is observed in all our GaAs
(100) 2D hole samples, including the square well sample which
shows an FT spectrum qualitatively similar to
Fig.~\ref{fig:fft}(b). In the remainder of the paper we
concentrate on the data from the heterojunction sample as it shows
a much higher spin splitting than the square well.

We first compare the measured FT spectrum of the SdH oscillations
with that of calculated magneto-oscillations of the density of
states (DOS) at the Fermi energy ($E_F$) \cite{Winkler00,
Winkler01}. We obtain the Landau fan chart at $B>0$ by evaluating
an $8 \times 8$ $\mathbf{k} \cdot \mathbf{p}$ Hamiltonian that
fully takes into account SO coupling due to both the structure
inversion asymmetry of the GaAs/AlGaAs heterojunction as well as
the bulk inversion asymmetry of the underlying zinc blende
structure \cite{Winkler93,Trebin}. In our calculations we assumed
that the system formed an accumulation layer. For such systems
Stern \cite{Stern74} pointed out that the Hartree potential
depends on the concentration of minority impurities. We assumed
that the concentration of unintentional minority impurities in the
GaAs layer was $1 \times 10^{14}$~cm$^{-3}$ and we used a spacer
width of $25$~nm; both assumptions are consistent with our sample
parameters. We note that one should not expect a one-to-one
correspondence between the measured SdH oscillations and the
calculated magneto-oscillations of the DOS at $E_F$. For example,
the amplitude of the oscillations which determine the amplitudes
of the FT spectrum peaks are different from the SdH oscillations.
However, the peak positions in the FT spectrum -- the quantities
we are interested in here-- are not affected by these details
\cite{Winkler00, Winkler01}.

Figure~\ref{fig:fft}(c) shows the FT spectrum of the calculated
DOS magneto-oscillations for the sample parameters of
Figs.~\ref{fig:fft}(a) and (b), i.e., a heterojunction sample with
$p = 2.3 \times 10^{11}$~cm$^{-2}$. The spectrum shows three main
peaks, marked by $f_-$, $f_+$, and $f_\mathrm{tot}$. The positions
of these peaks are in good agreement with the corresponding three
peaks in the experimental data [Fig.~\ref{fig:fft}(b)]. The
$f_\mathrm{avg}$ peak is also seen, albeit with a smaller
amplitude, in the FT spectrum of the calculated
magneto-oscillations. In Fig.~\ref{fig:fft}(c) we have indicated,
by open circles, the positions of FT peaks expected from the
relation $f = (h/e) p_\pm$, where $p_\pm$ are the densities of the
two spin subband densities calculated at $B = 0$ \cite{Winkler03}.
It is evident that the frequencies $f_-$ and $f_+$ do not directly
give the spin subband densities via the relation $p_\pm = (e/h)
f_\pm$. This anomalous behavior, like the unexpected peak at
$f_\mathrm{avg}$, reflects the fact that the spin precession along
the cyclotron orbits becomes nonadiabatic if the SO coupling in
some parts of $\mathbf{k}_\|$ space is sufficiently weak
\cite{Keppeler02}. Qualitatively, one can argue that the peak at
$f_\mathrm{avg}$ is due to the presence of cyclotron orbits in
which the holes move half the time along the constant energy
contour of the HHh band and then switch to the HHl contour
\cite{Winkler00}. This simplified picture cannot explain, however,
that the peaks at $f_\pm$ do not obey the relation $f_\pm = (h/e)
p_\pm$. We remark that anomalous magneto-oscillations were
observed also in GaAs (311)A 2D electrons and
holes~\cite{Winkler00, Keppeler02}.

We now concentrate on our results for the effective masses of the
HHl and HHh spin subbands.  In a 2D system with only one spin
subband occupied, or with degenerate subbands, the variation in
the amplitude $\Delta R_{xx}$ of the SdH oscillations with
temperature $T$ is commonly used to determine the carrier
effective mass $m^\ast$. This is done by fitting $\Delta R_{xx}$
to the Dingle factor, $\xi/\sinh\xi$, where $\xi\equiv 2 \pi^2
k_\mathrm{B} T / (\hbar\omega_c)$ and $\omega_c = eB/m^\ast$
\cite{Dingle, Adams}. Here $m^\ast$ is the fitting parameter. In
our system however, due to the spin splitting of the subbands
giving rise to beating patterns in the SdH oscillations, the two
masses cannot be deduced directly by fitting $\Delta R_{xx}$ to a
Dingle factor. In the SdH experiments performed by Eisenstein {\em
et al.}\ \cite{Eisenstein} on GaAs (100) 2D holes, the authors
assigned the low field oscillations (with frequency $f_-$) to the
SdH effect of the spin subband with the lighter mass, and the
oscillations at higher fields (with frequency $f_\mathrm{tot}$) to
the total 2D hole density. They then determined the mass for HHl,
$m_-^\ast$, by fitting the temperature dependence of the low field
oscillations to a Dingle factor. To deduce the HHh mass,
$m_+^\ast$, they assumed that the two spin subbands have parabolic
dispersion curves $E_\pm (\mathbf{k}_\|) = \hbar^2 k_\|^2 /
(2m_\pm^\ast)$, where $\mathbf{k}_\|$ is the in-plane wave vector
and $m^\ast_\pm$ are constants independent of the energy $E$. This
model implies the relation $m_-^\ast / m_+^\ast = f_- /
(f_\mathrm{tot} - f_-)$ which was used in Ref.~\cite{Eisenstein}
to obtain $m_+^\ast$.

\begin{figure}
\centering
\includegraphics{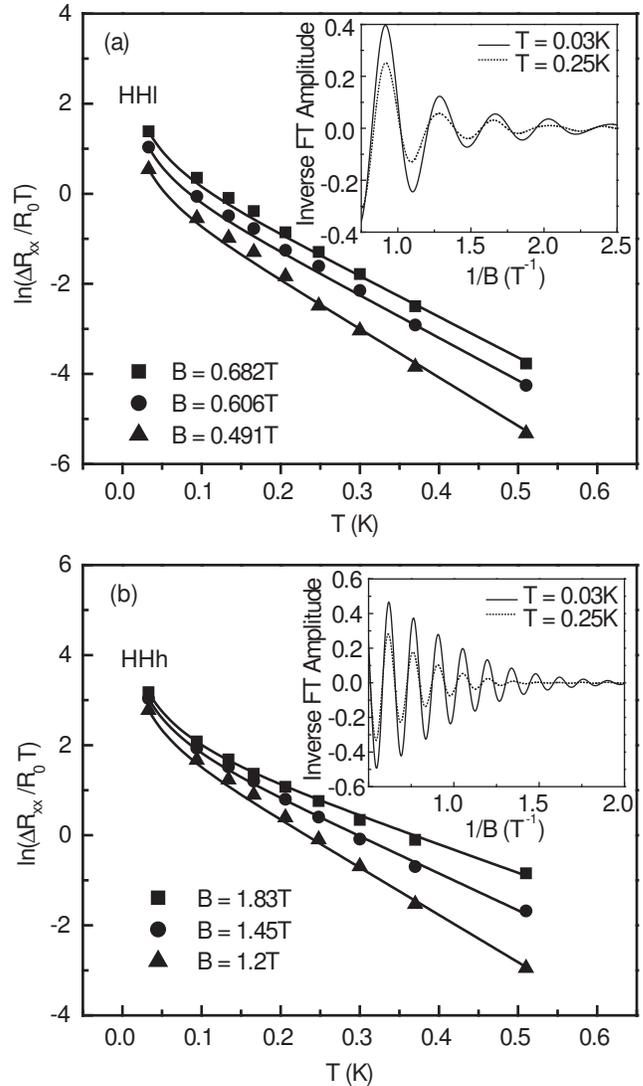}
\caption{\label{fig:ifft}Variation of $\Delta R_{xx}$ with
temperature and its fit to the Dingle factor with $m^\ast$ as the
fitting parameter for (a) the HHl and (b) the HHh subbands,
respectively. Insets in (a) and (b): After isolating the peaks
$f_-$ and $f_+$ by square windows [see Fig~\ref{fig:fft}(b)],
inverse FT is performed on these peaks separately to deduce SdH
oscillations of the HHl and HHh subbands.}
\end{figure}

\begin{figure}
\centering
\includegraphics{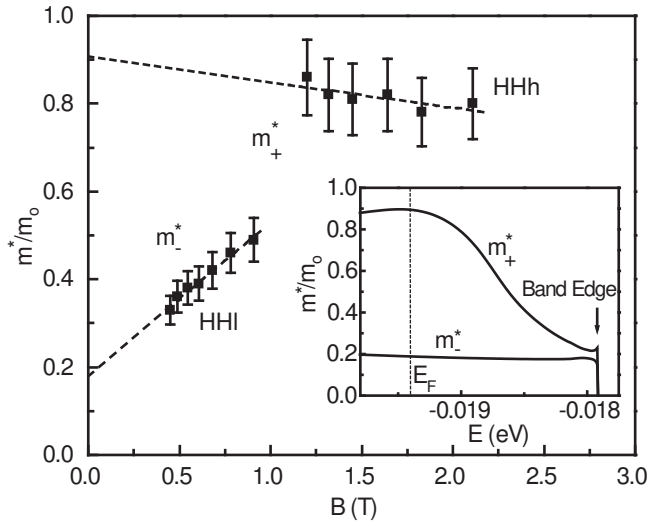}
\caption{\label{fig:mvsb}Values for $m_-^\ast$ and $m_+^\ast$ for
the HHl and HHh subbands, deduced from fits of SdH oscillation
amplitudes to the Dingle factor for the heterojunction sample,
with $p=2.3\times10^{11}$~cm$^{-2}$ ($m_0$ is the free electron
mass). Inset: $m^\ast$ calculated from the subband structure at
$B=0$.}
\end{figure}

In our study, the FT spectrum of the oscillations allows us to
independently determine $m^\ast_\pm$ for the two spin subbands. We
isolated the individual peaks, $f_-$ and $f_+$ for HHl and HHh,
with square windows [Fig.~\ref{fig:fft}(b)] and calculated their
inverse FT \cite{IFT}. The insets in Figs.~\ref{fig:ifft}(a) and
(b) show the inverse transform for each of these peaks. The SdH
oscillations in these insets can then be attributed to the HHl and
HHh spin subbands, respectively. The main parts of
Fig.~\ref{fig:ifft} show the temperature variation of the
amplitude of these oscillations. We fitted this variation at
particular $B$ values to a Dingle factor for each subband
[Figs.~\ref{fig:ifft}(a) and (b)] and deduced the $m^\ast$ values
as the fitting parameters. Figure~\ref{fig:mvsb} shows $m^\ast$
values determined as a function of $B$ for a hole density of
$2.3\times10^{11}$~cm$^{-2}$. The amplitude variation of the
oscillations fitted to the Dingle factor at different values of
$B$ should in principle give the same mass. In our system however,
$m_-^\ast$ has a strong, nearly linear dependence on the value of
$B$ at which the variation of $R_{xx}$ with temperature is
analyzed. Only $m_+^\ast$ is approximately independent of $B$. The
origin of the field dependence of $m_-^\ast$ which was also seen
in Ref.~\cite{Eisenstein}, remains unknown. Linear extrapolation
of the data to $B = 0$ (dashed lines in Fig.~\ref{fig:mvsb}),
suggests $m_+^\ast \simeq 0.9$~$m_0$ and $m_-^\ast \simeq
0.2$~$m_0$, where $m_0$ is the mass of free electrons. Note that
the ratio of the hole densities of the two subbands is not equal
to the ratio of their masses. Hence, in general it cannot be
assumed that the bands are parabolic and, in particular,
$m_+^\ast$ cannot be deduced from the ratio of the spin subband
densities and $m_-^\ast$.

We determined $m^\ast$ at different hole densities ranging from
2.0 to $2.4 \times 10^{11}$~cm$^{-2}$ for the heterojunction
sample, where we used front and back gate biases to vary the
density. The qualitative trends were similar in all cases. The
strong dependence of $m_-^\ast$ on $B$ was observed at all
densities and $m_+^\ast$ was found to be relatively independent of
$B$. Quantitatively, $m_+^\ast$ ranged from 0.75 to 0.91 and
$m_-^\ast$ extrapolated to a $B=0$ value between 0.18 to 0.25 in
this density range.

%%%%%%%%%%%%%%%%%%%%%%%%%%%%%%%%%%%%%%%%%%%%%%%%%%%%%%%%%%%%%%%%%%
% \newpage
% \section{Discussion and Conclusion}

We compare the measured effective masses with an energy-dependent
DOS effective mass
\begin{equation}
\label{eq:dos_emass}
 \frac{m^\ast_\pm (E)}{m_0} =
 \frac{1}{\pi} \, \frac{\hbar^2}{2m_0}
 \int \!d^2 k_\| \, \delta \big[ E - E_\pm
 (\mathbf{k}_\|) \big]
\end{equation}
for the two spin subbands. Equation (\ref{eq:dos_emass}) gives the
effective mass $m^\ast_\pm (E)$ one needs locally at energy $E$ to
approximate $E_\pm (\mathbf{k}_\|)$ by a parabolic dispersion
which yields the same DOS like $E_\pm (\mathbf{k}_\|)$. Our
numerical calculations of $E_\pm (\mathbf{k}_\|)$ follow
Ref.~\cite{Winkler93}. We then evaluate Eq.\ (\ref{eq:dos_emass})
by means of analytical quadratic Brillouin zone integration
\cite{Winkler93a}. We remark that the SdH oscillations are an
effect taking place at $E_F$. Therefore, the DOS $m^\ast$, defined
by Eq. (\ref{eq:dos_emass}) is the appropriate quantity for
comparison with the effective mass deduced experimentally from the
temperature dependence of the SdH oscillations. However, one
should not expect to find a one-to-one correspondence between
experiment and theory because the temperature dependence of
magneto-oscillations in 2D systems can be more complicated than
implicitly assumed by the Dingle formula \cite{Fang77,Champel}.

The inset in Fig.~\ref{fig:mvsb} shows the DOS effective mass
calculated for $p = 2.3 \times 10^{11}$~cm$^{-2}$. The subband
edge is at $-17.9$~meV and $E_F$ at $-19.4$~meV. We obtain an
effective mass at $E_F$ of 0.9~$m_0$ and 0.2~$m_0$ for the HHh and
HHl bands, respectively. This is in good agreement with our
experimental results (Fig.~\ref{fig:mvsb}). From the curve for the
heavier mass, we see that the HHh band is highly nonparabolic. We
note that symmetry requires that in a Taylor expansion of the
subband dispersion $E_\pm(\mathbf{k}_\|)$, the $B=0$ spin
splitting is characterized by terms proportional to odd powers of
$\mathbf{k}_\|$. Only the spin-independent terms in the Taylor
expansion are proportional to even powers of $\mathbf{k}_\|$
\cite{odd}. In 2D \emph{hole} systems the odd-power terms give
rise to a spin splitting that is comparable in magnitude to the
spin-independent part of the subband dispersion \cite{Broido,
Ando85, Ekenberg, Bangert}. The spin-split subbands thus have
nonparabolic dispersion curves with strongly spin-dependent
masses. We remark that spin-split \emph{electron} subbands can
usually be characterized by the same value of $m^\ast$ because in
the electron case, spin splitting represents only a small
correction to the spin-independent part of the subband dispersion
\cite{Winkler93}.

In conclusion, we have measured SdH oscillations in GaAs (100) 2D
holes. Through Fourier analysis of the oscillations, we determined
the splitting as well as the effective masses of the two spin
subbands. The results are in good agreement with the calculations.
There remain however, some open questions regarding the origin of
the average peak observed in the FT spectrum and the dependence of
experimentally determined mass values on $B$, especially for
$m_-^\ast$.

We thank the DOE, ARO and NSF for support.

%%%%%%%%%%%%%%%%%%%%%%%%%%%%%%%%%%%%%%%%%%%%%%%%%%%%%%%%%%%%%%%%%%


\begin{thebibliography}{99}
\bibitem{IFT}
For the $f_-$ peak, the low frequency shoulder can be included or
ignored when applying the square window. The inverse FT showed a
slight dependence on where exactly the window was applied.
However, the portion of the inverse FT which exhibited this
dependency was not used in further analysis.
\bibitem{odd} In inversion asymmetric systems we can also have
 spin-independent terms proportional to odd powers of $k_\|$.
 However, it is not possible to have spin-dependent terms
 proportional to even powers of $k_\|$.
\end{thebibliography}
\end{document}